\documentclass{aa}
\usepackage{psfig}
\usepackage{graphicx}
\begin{document}
   \headnote{Letter to the Editor}

   \title{The mysterious eruption of V838 Mon}

   \author{
           U. Munari\inst{1}
           \and
           A.Henden\inst{2}
	   \and
           S. Kiyota\inst{3}
	   \and
	   D. Laney\inst{4}
	   \and
           F. Marang\inst{4}
	   \and
	   T. Zwitter\inst{5}
	   \and
	   R.L.M. Corradi\inst{6}
	   \and
	   S. Desidera\inst{1}
	   \and
	   P. Marrese\inst{1}
	   \and
	   E. Giro\inst{1}
           \and
	   F. Boschi\inst{1}
           \and
	   M.B. Schwartz\inst{7}
          }

   \offprints{U.Munari \email{munari@pd.astro.it}}

   \institute{
INAF-Osservatorio Astronomico di Padova, Sede di Asiago,
I-36012 Asiago (VI), Italy
\and
Univ. Space Research Ass./U. S. Naval Observatory,
P. O. Box 1149, Flagstaff AZ 86002-1149, USA
\and
VSOLJ, 1-401-810 Azuma, Tsukuba 305-0031 Japan
\and
South African Astronomical Observatory, P.O.Box 9, Observatory 7935, 
South Africa
\and
University of Ljubljana, Department of Physics, Jadranska 19, 
1000 Ljubljana, Slovenia
\and
Isaac Newton Group of Telescopes, Apartado de Correos 321, 38700 Santa
Cruz de La Palma, Canarias, Spain
\and
Tenagra Observatory, HC2 Box 292 Nogales, AZ  85621, USA
             }

   \date{Received ... / Accepted ...}

\abstract{V838~Mon is marking one of the most mysterious stellar outbursts 
on record. The spectral energy distribution of the progenitor resembles an
under-luminous F main sequence star (at $V$=15.6 mag), that erupted into a
cool supergiant following a complex and multi-maxima lightcurve (peaking at
$V$=6.7 mag). The outburst spectrum show BaII, LiI and lines of several
$s-$elements, with wide P-Cyg profiles and a moderate and retracing emission
in the Balmer lines. A light-echo discovered expanding around the object
helped to constrain the distance ($d$=790$\pm$30 pc), providing $M_V=+4.45$
in quiescence and $M_V=-4.35$ at optical maximum (somewhat dependent on the
still uncertain E$_{B-V}$=0.5 reddening). The general outburst trend is
toward lower temperatures and larger luminosities, and continuing so at the
time of writing. The object properties conflict with a classification within
already existing categories: the progenitor was not on a post-AGB track and
thus the similarities with the born-again AGB stars FG~Sge, V605~Aql and
Sakurai's object are limited to the cool giant spectrum at maximum; the cool
spectrum, the moderate wind velocity (500 km~sec$^{-1}$ and progressively
reducing) and the monotonic decreasing of the low ionization condition
argues against a classical nova scenario. The closest similarity is with a
star that erupted into an M-type supergiant discovered in M31 by Rich et al.
(1989), that became however much brighter by peaking at $M_V=-9.95$, and with 
V4332~Sgr that too erupted into an M-type giant (Martini et al. 1999) and that
attained a lower luminosity, closer to that of V838~Mon. M31-RedVar, V4332~Sgr
and V838~Mon could be all manifestations of a new class of astronomical objects.

\keywords{ Stars: supergiants - Stars: novae - Stars: individual: 
           V838 Mon - Stars: mass-loss - 
           ISM: jets and outflows}}

   \maketitle

\section{Introduction}

The previously unnoticed, highly peculiar object V838~Mon was discovered in
outburst by Brown (2002) on January 6. The complex lightcurve, cool colors
at maximum (in spite of a $\bigtriangleup m = 9$ mag amplitude), strong mass
loss and a spectrum rich in $s-$process elements, a peak $V=6.7$ mag and a
favorable position of the celestial equator favored massive world-wide
interest and an observational effort that has so far resulted in 23 IAU
Circulars.
In this {\sl Letter} we present and discuss our astrometric, photometric,
spectroscopic, imaging and polarimetric observations. The basic properties
of V848~Mon in quiescence and outburst are derived, and its nature outlined.
Only a fast and preliminary analysis of the large amount of gathered
information will be possible in this {\sl Letter}. A more complete data
analysis will be performed elsewhere.

\begin{figure*}[!t]
\centering
\includegraphics[width=18cm]{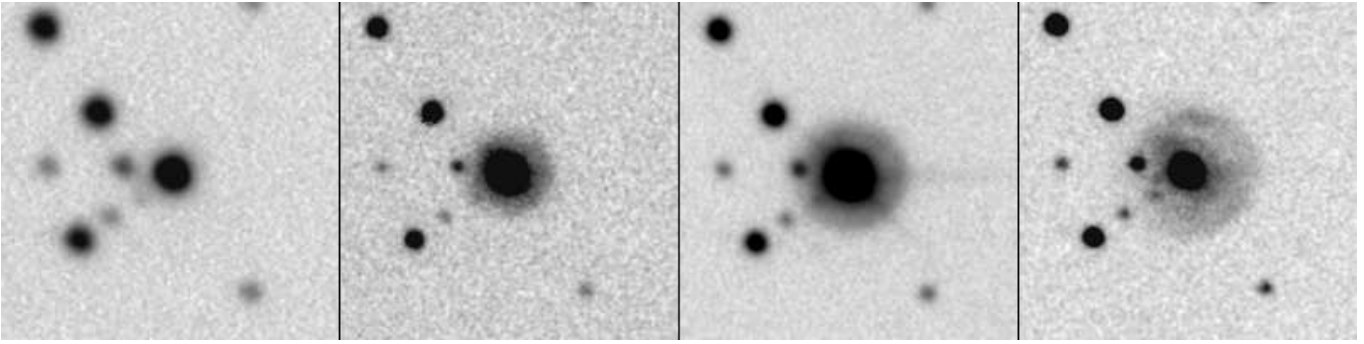}
      \caption{Expansion of the light-echo around V838~Mon, revealing a
               previously invisible ring of circumstellar material. $U$ band
               67$\times$67 arcsec images obtained with the USNO 1m
               telescope (North to top, East to the left). 
               Dates (seeing in arcsec, $U$ mag of central
               V838~Mon) from left to right: Jan 13 (3.2", $U$=13.33), Feb
               27 (2.3", $U$=12.05), March 10 (2.5", $U$=10.62) and March 27
               (2.2", $U$=12.28).}
\label{ring}
\end{figure*}

We obtained CCD {\sl UBV$R_c$I$_c$} photometry of V838~Mon with the USNO 1m
telescope in Flagstaff and a privately operated 25~cm telescope in Tsukuba,
Japan, against a photometric sequence that we calibrated with respect to
Landolt's equatorial standards (sequence and identification chart available
via http://ulisse.pd.astro.it/V838\_Mon/). The photometry is reported in
Tables~1 and 2 (with errors a few units of the last decimal figure). $U$
band imaging has also been provided by the 0.81m telescope of Tanagra Obs.
(Arizona) and the NOT 2.5m and WHT 4.2m telescopes in La Palma (Canary
Islands).
{\sl JHKL} photometry (on the Carter system) of V838~Mon has been secured
with the 0.75-m telescope at SAAO equipped with the IRP Mk~II photometer.
The magnitudes are listed in Table~3, with errors around 0.01 mag.
High resolution spectra of V838~Mon have been regularly obtained since
outburst onset with the Echelle+CCD spectrograph of the 1.82 m Asiago
telescope, covering the range 4600--9800 \AA\ at 18500 resolving power.
Finally, medium resolution spectroscopy (res.power 6000) has been obtained
with AFOSC at the 1.82 m Asiago telescope over the 3900-7700 \AA\ range, as
well as spectropolarimetry (2 and 4 \AA/pix) and polarimetric imaging
($UBVR_cI_c$) on several nights each month since the onset of the outburst.

\begin{table}
\centering
\caption[]{USNO {\sl UBVRI} photometry of V838 Mon in outburst}
\begin{tabular}{cccccc}
\hline
 HJD      &  $V$  & $B-V$ & $U-B$ & $V-R_{\rm c}$ &  $R-I_{\rm c}$ \\    
\\
2452285.7231 & 9.977 & 1.756 &1.917 &0.959 &0.942 \\
2452287.8159 & 9.881 & 1.687 &1.760 &0.925 &0.889 \\
2452288.8061 & 9.896 & 1.681 &1.771 &0.938 &0.873 \\
2452309.6288 & 7.473 & 1.130 &0.261 &0.717 &0.804 \\
2452312.6482 & 6.926 & 0.981 &0.239 &      &      \\
2452313.6567 & 7.039 & 1.107 &0.398 &      &      \\
2452314.7245 & 7.258 & 1.128 &0.508 &0.699 &0.759 \\
2452315.6028 & 7.420 & 1.156 &0.570 &0.703 &0.795 \\
2452316.6553 & 7.614 & 1.191 &0.662 &0.721 &0.731 \\
2452317.6014 & 7.727 & 1.254 &0.689 &0.761 &0.806 \\
2452321.6405 & 7.913 & 1.494 &      &      &      \\
2452322.7572 & 7.896 & 1.654 &1.025 &0.942 &0.862 \\
2452330.6609 & 8.124 & 2.016 &1.503 &1.079 &1.037 \\
2452332.6254 & 8.186 & 2.137 &1.726 &1.169 &0.987 \\
2452333.6359 & 8.140 & 2.148 &1.764 &1.150 &1.002 \\
2452337.6832 & 7.624 & 1.811 &1.324 &0.997 &0.962 \\
2452338.6850 & 7.506 & 1.753 &1.301 &0.976 &0.901 \\
2452342.6248 & 7.205 & 1.781 &1.539 &0.996 &0.906 \\
2452343.6502 & 7.195 & 1.790 &1.631 &0.974 &0.935 \\
2452344.6465 & 7.164 & 1.785 &1.661 &0.978 &0.900 \\
2452360.7085 & 7.622 & 2.366 &2.295 &      &      \\ 
2452365.6932 & 7.735 & 2.520 &2.771 &1.360 &1.053 \\
2452369.6217 & 7.943 & 2.595 &2.590 &      &      \\
2452373.6287 & 8.254 & 2.635 &2.555 &1.418 &1.315 \\ 
\hline
\end{tabular}
\label{tab1}
\end{table}

\begin{table}
\centering
\caption[]{Tsukuba {\sl BVI} photometry of V838 Mon in outburst}
\begin{tabular}{crrr|crrr}
\hline
 HJD      &  $B$~  & $V$~ & $I_c$~ &  HJD   &  $B$~  & $V$~ & $I_c$~ \\
&&&&&&\\
283.998 & 12.16 &  10.10 &  8.28 &  312.902 &  8.09 &   6.80 &  5.38 \\
285.092 & 11.95 &   9.93 &  8.03 &  316.030 &  8.79 &   7.44 &  5.94 \\
286.023 & 11.77 &   9.79 &  7.93 &  317.040 &  8.95 &   7.61 &  6.06 \\
287.098 & 11.77 &   9.79 &  7.96 &  317.956 &  9.09 &   7.66 &  6.06 \\
290.035 & 11.76 &   9.83 &  7.98 &  324.025 &  9.72 &   7.83 &  5.93 \\
292.952 & 11.95 &   9.95 &  8.01 &  324.959 &  9.77 &   7.83 &  5.93 \\
293.984 & 12.06 &  10.01 &  8.06 &  325.940 &  9.88 &   7.88 &  5.98 \\
297.060 & 12.06 &  10.03 &  8.08 &  337.927 &  9.44 &   7.54 &  5.61 \\
298.005 & 12.06 &  10.02 &  8.08 &  341.056 &  9.16 &   7.12 &  5.26 \\
299.009 & 12.14 &  10.07 &  8.15 &  341.949 &  9.01 &   6.95 &  5.16 \\
300.000 & 12.19 &  10.12 &  8.13 &  343.044 &  8.91 &   7.02 &  5.14 \\
301.931 & 12.35 &  10.25 &  8.22 &  343.931 &  8.96 &   6.94 &  5.14 \\
304.011 & 12.64 &  10.48 &  8.39 &  344.980 &  9.19 &   6.99 &  5.13 \\
304.938 & 12.73 &  10.56 &  8.44 &  346.950 &  9.17 &   7.09 &  5.20 \\
306.987 & 12.85 &  10.67 &  8.51 &  349.007 &  9.45 &   7.25 &  5.28 \\
309.904 &  8.55 &   7.29 &  5.76 &  350.024 &  9.57 &   7.26 &  5.24 \\                                   
311.906 &  7.99 &   6.66 &  5.24 &  354.011 &  9.84 &   7.39 &  5.26 \\
\hline
\end{tabular}
\label{tab2}
\end{table}

\section{The light-echo and distance to V838 Mon}

A nebula around V838~Mon has been discovered and seen to ``{\sl expand}"
with time on broad-band monitoring images obtained with the USNO 1m
telescope, a sample of which are shown in Figure~1. The nebula is best seen
in $U$ band and it is progressively less evident at longer wavelengths, a
pattern typical of light scattering. We interpret the expanding nebula as a
light echo of the outburst produced by circumstellar material lost by the
progenitor. The nebula is just reflecting light from the central star and it
is not self-emitting, because [OIII] imaging and long slit spectroscopy
failed to reveal emission lines from it. 

\begin{figure}[!h]
\centering
\includegraphics[width=8.7cm]{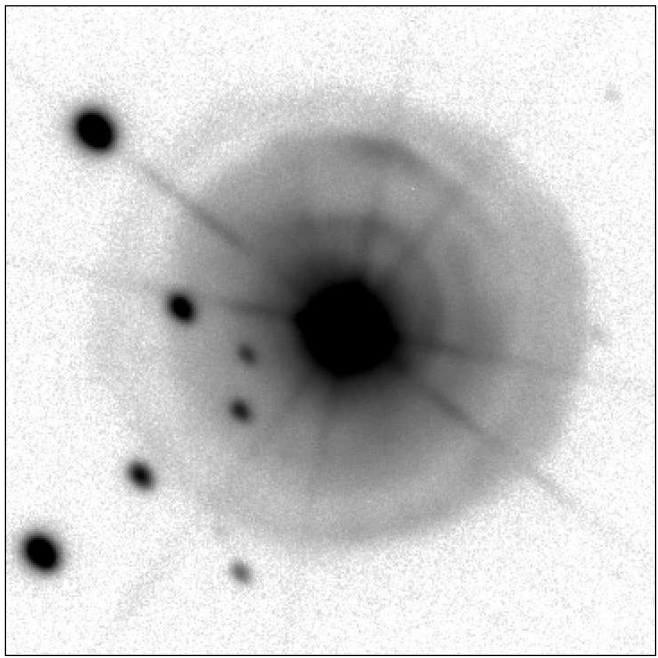}
      \caption{A median averaged combination of seven 30-sec $U-$band images
               obtained with the 4.2m WHT telescope on March 28, 2002
               showing the light-echo (image size 40 arcsec; North to top,
               East to the left; seeing 0.9 arcsec,
               $U$=12.65 mag for central V838~Mon).}
\label{wht}
\end{figure}

\begin{table}
\centering
\caption[]{SAAO {\sl JHKL} photometry of V838 Mon in outburst}
\begin{tabular}{ccccc}
\hline
 HJD      &  $J$  & $H$ & $K$ & $L$ \\
\\
2452285.442 & 6.828 & 6.202 & 5.971 & 5.522 \\
2452309.284 & 4.947 & 4.369 & 4.041 & 3.654 \\
2452309.502 & 4.905 & 4.331 & 4.005 & 3.600 \\
2452311.278 & 4.383 & 3.834 & 3.512 & 3.132 \\
2452313.288 & 4.495 & 3.963 & 3.645 & 3.242 \\
2452314.291 & 4.712 & 4.178 & 3.868 & 3.475 \\
2452315.292 & 4.881 & 4.364 & 4.071 & 3.727 \\
2452316.412 & 5.007 & 4.475 & 4.190 & 3.875 \\
2452317.384 & 5.061 & 4.509 & 4.230 & 3.885 \\
2452318.356 & 5.053 & 4.487 & 4.173 & 3.885 \\
2452319.349 & 5.037 & 4.408 & 4.116 & 3.825 \\
2452320.429 & 4.951 & 4.323 & 4.030 & 3.688 \\
2452321.363 & 4.919 & 4.259 & 3.971 & 3.682 \\
2452323.356 & 4.803 & 4.136 & 3.837 & 3.572 \\
2452324.336 & 4.762 & 4.086 & 3.774 & 3.506 \\
\hline
\end{tabular}
\label{tab3}
\end{table}

The light-echo expansion rate is 0.44$\pm$0.017 arcsec day$^{-1}$ in
diameter, reaching 30 arcsec by early April on the USNO 1m $U$ images used
to monitor its development. The light producing the echo is from the second,
brighter maximum in the lightcurve (see sect.5). Assuming a spherical
symmetric distribution of the circumstellar dust, the expansion rate sets
the distance of V838~Mon to 790$\pm$30 pc. A higher resolution image in $U$
band obtained with the WHT 4.2m telescope at La Palma (Canary Islands) for
March 28 is presented in Figure~2. This much finer image does not support
our first impression (Henden et al. 2002) of a central hole in the nebula
based on the USNO 1m telescope lower resolution images and the assumption
that the scattered light was from the first maximum in the lightcurve. The
outburst light sweeping through the circumstellar material allows us to read
the recent mass loss history of the AGB progenitor: assuming a typical 15
km~sec$^{-1}$ velocity for its wind, the light-echo has reached by early
April material lost $\sim$3750 years ago. Some circularly symmetric
brightness enhancements are evident, reminiscent of similar structure seen
in the surface brightness distribution of several planetary nebulae
(cf. Balick et al. 2001, their Fig.1), usually
interpreted as variations in the progenitor AGB mass loss rate. Their
angular separation indicate a $\sim$1200 year recurrence time.

Could the light-echo be originating instead in a slab of interstellar dust?
We believe not for three independent reasons. First, given the perfect round
shape of the echo, the slab should be flat and perpendicular to the line of
sight, i.e. perpendicular to the galactic plane and therefore hardly a
stable dynamical condition. Second, the concentric structures seen in better
detail in Figure~2 do not expand with the light-echo and instead seem to
remain fixed in space. They are therefore real density structures. The very
low probability that such concentric structures develop in the interstellar
space and that they also lie perfectly aligned with the line of sight to
V838~Mon argue strongly in favor of a circumstellar origin of the
light-echo. Finally, V838~Mon progenitor was detected by IRAS (source
07015-0346) in the 60 and 100~$\mu$m bands with fluxes of 1.4 and 4.6 Jansky
($\pm$10\%), respectively, indicating dust emission at low temperature.

\section{The reddening and polarization}

Zwitter and Munari (2002) on high resolution spectra taken in January, 
identified narrow components of NaI and KI superimposed on the wider P-Cyg stellar
profiles that if interpreted as interstellar would have consistently
indicated a $E_{B-V}$=0.80$\pm$0.05 reddening when their equivalent width is
compared with the Munari and Zwitter (1997) calibration. Figure~3 shows the
narrow components as they appeared in late March.  On the Neckel and Klare
(1980) extinction maps for the V838~Mon region, an interstellar
$E_{B-V}\sim$0.25 is reached at the 790 pc distance to the object, while
$E_{B-V}$=0.80 pertains to distances larger than 3~kpc. Therefore,
contribution from circumstellar material must be invoked to explain the
intensity of narrow NaI and KI components. How much this circumstellar
material also contributes to the reddening is however unknown because, for
example, the dust-to-gas ratio could largely differ from the typical
interstellar value or the reddening law be different from the standard
$R_V$=3.1 one.

The reddening affecting V838 Mon probably lies between a minimum
$E_{B-V}\sim$0.25 implied by the distance and a maximum that could be
identified with the $E_{B-V}$=0.80 derived from the intensity of narrow NaI
and KI components. A midpoint $E_{B-V}$=0.5 value will be adopted in the rest
of this {\sl Letter}. $E_{B-V}$=0.5 is supported by polarimetry. We have
repeatedly measured during February and March the polarization of V838~Mon
both in $UBVR_cI_c$ bands and over 4400-7900 \AA\ medium resolution spectra
with AFOSC in polarimetric mode attached to the Asiago 1.82m telescope.  The
data indicate a polarization constant in time, with a wavelength slope
characteristic of an interstellar origin (Serkowski's law), amounting to
2.6\% at 5500 \AA\ at a position angle 150$^\circ \pm 2$. Adopting the
average relation $p(\%,V)/E(B-V)=5.0$ between polarization and reddening
found by Serkowski et al. (1975), the observed $p=$2.6\% corresponds to
$E_{B-V}$=0.52.

\section{The progenitor}

An accurate astrometric position for V838~Mon in outburst has been obtained
from USNO 1m images (linked to the USNO-A2.0 local grid). It allows an
identification of the progenitor with an anonymous $V\sim$15.6 mag star that
has positional measurements in both the USNO A2.0 and 2MASS catalogues:
\begin{center}
\begin{tabular}{cccl}
$\alpha_{J2000}$&$\delta_{J2000}$&epoch&\\
07:04:04.81 & --03:50:50.9 & 2002 & ours\\
07:04:04.82 & --03:50:50.5 & 1997 & 2MASS\\
07:04:04.85 & --03:50:51.1 & 1953 & POSS-I\\ 
\end{tabular}
\end{center}

The marginal or null proper motion supports a low space velocity of the
progenitor, and thus a probable partnership with galactic disk stars (a
100~km~sec$^{-1}$ transverse velocity at 790 pc corresponds to a 1.3 arcsec
displacement during the time elapsed between POSS-I and our observations). 
At galactic coordinates $l=217.80$ $b=+1.05$, the height over the galactic 
plane is just $z=$13 pc, again supporting a link to Pop I stars. The radial
velocity of the progenitor is unknown, and values derived from absorption
lines in outburst are severely affected by the P-Cyg profiles of the lines
(even using LaII lines that rank among the sharper ones with marginal
emission components, the mean heliocentric velocity for late January was
+54.4$\pm$0.8, for late February $-84.4\pm 0.2$ and for late March $-57.5\pm
0.5$ km~sec$^{-1}$).

The brightness of the progenitor has been measured on Palomar and SERC
plates by comparison with the photometric sequence we have calibrated:
\begin{center}
\begin{tabular}{cccccc}
Plate  &  Date     & UT    &band& mag   &          \\
so0662 & 1953.0445 & 06.42 & B     &16.00  & POSS-I   \\
se0662 & 1953.0445 & 07.52 &~~R$_c$&15.35  &   "      \\
sb0772 & 1983.0431 & 11.90 & B     &16.10  & SERC     \\
sr0772 & 1989.1732 & 10.65 &~~R$_c$&15.30  &   "      \\
\end{tabular}
\end{center}

\begin{figure}
\centering
\includegraphics[width=8.7cm]{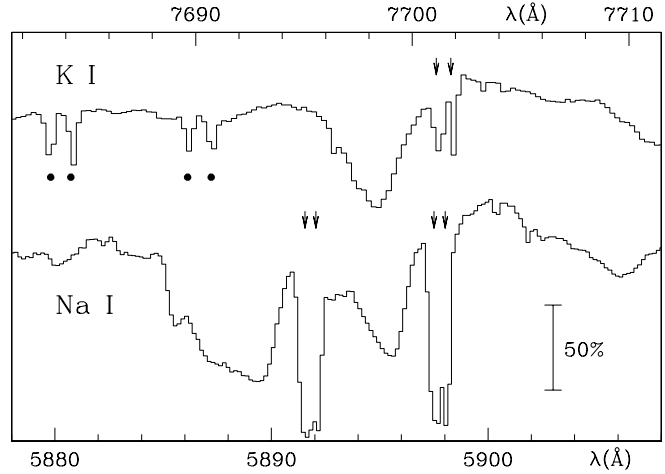}
      \caption{The spectrum of V838~Mon around NaI and KI lines for March 29,
               2002.  The dots mark telluric absorptions, the arrows the
               narrow components discussed in the text.
               The wide absorptions are the stellar NaI and KI profiles.
               The bar marks the intensity scale with respect to the 
               normalized continuum.}
\label{NaIKI}
\end{figure}

\begin{figure}
\centering
\includegraphics[width=8.7cm]{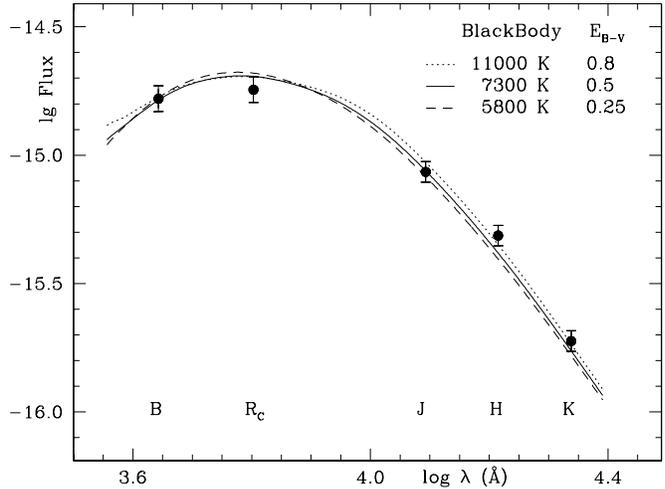}
      \caption{Spectral energy distribution of V838~Mon in quiescence,
               for the adopted $E_{B-V}$=0.5, and the extrema of the
               range of possible values discussed in sect.3}
\label{SEDq}
\end{figure}

The data support little or no variability of V838~Mon in quiescence during
the last half century. They can be therefore safely combined with the
near-IR 2MASS measurements ($J$=13.91, $H$=13.48 and $K$=13.35 obtained in
1997) to derive the spectral energy distribution of the progenitor, which is
done in Figure~4. Even the maximum allowed temperature in quiescence (11000
K) is not enough to ionize the circumstellar material that produced the
light-echo, and in fact a planetary nebula was not observed in quiescence. It
is worth to note that the galactic equator has been surveyed several times
from both hemispheres in search for emission line objects, and the progenitor
of V838~Mon has never been logged in as an emission source (in spite of the
favorable magnitude). Several emission line objects have been discovered by
such surveys in the region (for example AS~143 lies just 16 arcmin from
V838~Mon).

At $d$=790~pc, $E_{B-V}=$0.5 and M$_V$=+4.46, the best fitting 7300~K
blackbody corresponds to a progenitor of radius of $R\sim 0.8$ R$_\odot$ and
$L\sim 1.6$ L$_\odot$, 4.5$\times$ less luminous (1.63 mag) than a
corresponding F0~V main sequence star (cf. Figure~8).

\begin{figure}
\centering
\includegraphics[width=8.7cm]{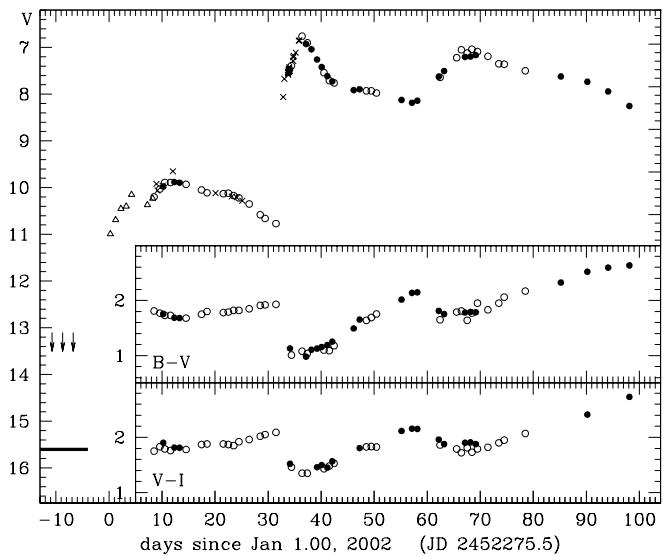}
      \caption{$V$, $B-V$ and $V-I_c$ lightcurves of the outburst of
               V838~Mon.  Dots mark data from Table~1, open circles from
               Table~2.  Crosses and open triangles are values from various
               IAUC and VSNET circulars (mainly from SAAO, D.West,
               P.Sobotka, L.Smelcer, F.Lomoz and J.Bedient). Arrows mark
               {\sl ``fainter than"} conditions from IAUC 7785, and the
               solid line indicates the quiescence brightness.}
\label{lightcurve}
\end{figure}

\begin{figure}
\centering
\includegraphics[width=8.7cm]{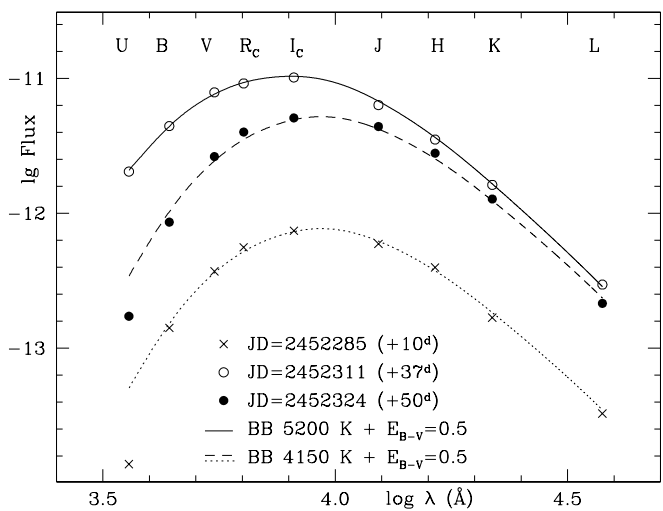}
      \caption{$UBVR_cI_cJHKL$ energy distribution of V838~Mon in outburst at
               three dates: first maximum, second maximum and mid plateau
               (data from Tables~1 and 3).}
\label{SEDo}
\end{figure}

\begin{figure}
\centering
\includegraphics[width=8.7cm]{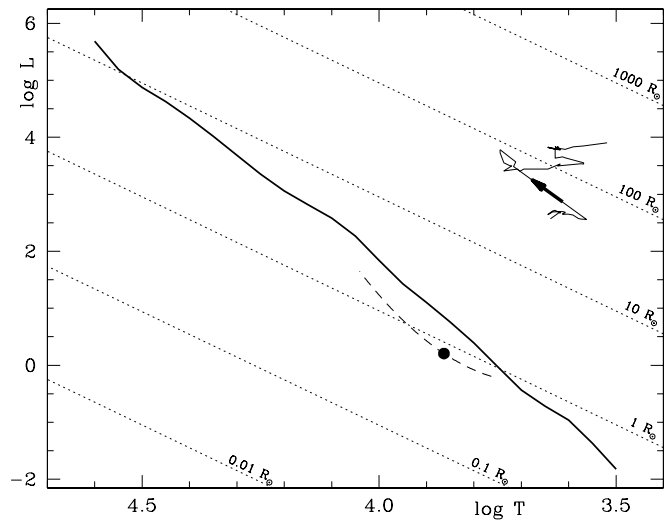}
       \caption{V838~Mon on the {\sl log L -- log T} digram. The diagonal
               thick line is the Main Sequence (O5 to M5, from Drilling and
               Landolt 2000). The solid circle gives the progenitor position
               and the curve traces the outburst path (from Jan 11 to Apr 1,
               direction arrowed) for the adopted $d=790$~pc and
               $E_{B-V}=$0.5. The range of locations for the progenitor (and
               equivalent displacements for the outburst path) are indicated
               by the dashed line (extrema at $d=790$ pc, $E_{B-V}=$0.25 and
               $d=8$~kpc, $E_{B-V}=$0.80)}
\label{HRdia}
\end{figure}

\section{The outburst}

V838~Mon went into outburst around the beginning of 2002, with a fast rise
to maximum (a week earlier it was still $V\geq$13.5 mag). The lightcurve up
to early April is presented in Figure~5. It is characterized by a complex
and rarely seen behavior.

A first maximum was reached by day +10 (see abscissae scale on Figure~5)
when the continuum energy distribution was characterized by a temperature of
4150~K (see dotted line in Figure~6), a second maximum at $+37^d$ peaked at
5200~K (solid line in Figure~6) and a third one at $+68^d$ reached 4600~K.
Decline from maxima are marked by monotonic cooling, with the last one
taking V838~Mon to 3400~K by day $+98^d$ (last point in Figure~5).

The outburst path on the {\sl logL - logT} diagram is presented in Figure~7.
It can be best described as a trend (with interruptions) toward the upper
right corner of the diagram, thus toward progressively cooler surface
temperatures and larger radii and greater luminosities.

The spectral appearance of V838~Mon in outburst matches the cool
photospheric temperatures indicated by the $UBVRI$ and $JHKL$ photometry.
Portions of high resolution sample spectra obtained with the Asiago Echelle
spectrograph are presented in Figure~8. The photospheric temperature at the
time these were taken was pretty similar, the $V-I$ index for the three dates
being +2.00, +2.14 and +2.22, respectively. The differences seen in the
intensity and profiles of the lines is therefore mainly affected by changing
wind structure and velocity, as well as lowering of surface gravity
(expanding radius).

\begin{figure*}[!ht]
\centering
\includegraphics[angle=270,width=18cm]{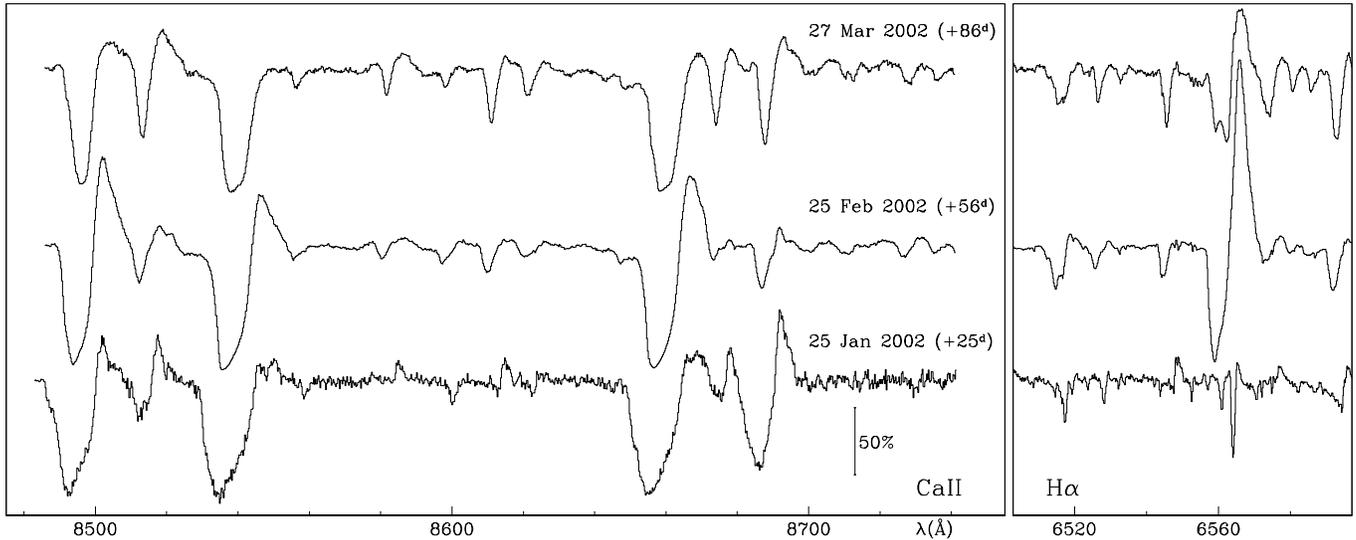}
      \caption{Small sections of sample Asiago Echelle spectra to 
               document the evolution around the near-IR Calcium triplet 
               and H$\alpha$. The bar marks the intensity scale with respect 
               to the normalized continuum.}
\label{CaII}
\end{figure*}

Late January spectra (toward the end of the decline from first maximum) are
characterized by wide P-Cyg profiles of species with low excitation
potential (E.P.). In particular CaII, BaII, NaI and LiI lines show
remarkably similar profiles and $-$500 km~sec$^{-1}$ terminal velocities.
Higher E.P. lines (like those of hydrogen and some FeI multiplets) show
instead almost pure absorptions, with narrow and Gaussian-like profiles
resembling those expected in a normal cool giant. In late February spectra
the P-Cyg terminal velocities of low E.P. species reduce (to $-$380
km~sec$^{-1}$) and those of higher E.P. widen, with all lines evolving
toward more homogeneous widths and shapes. The reduction of P-Cyg terminal
velocities for low E.P. species continued in late March spectra ($-$280
km~sec$^{-1}$), with an overall increase in the number and intensity of {\sl
normal} absorption lines. Balmer lines developed an emission component
around the second maximum, and have been declining since then.

No chemical abundance analysis will be attempted here, however an eye
inspection of the high resolution spectra reveals numerous and strong
absorptions from $s-$elements, with marked BaII and a LiI stronger than in
the FG~Sge and Sakurai's object spectra presented by Kipper (2001).

\section{The nature of V838 Mon}

The nature of V838 Mon is best described as a mysterious one. It is standard
in such cases to state that ``{\sl more observations are needed} ", which is
certainly true for V838~Mon.  Four key tasks for follow-up investigations
seem to be : ($a$) to better address the nature of the progenitor in
quiescence by reconstructing its photometric history from plate archives
around the globe (using the deep $UBVRI$ comparison sequence available from
http://ulisse.pd.astro.it/V838\_Mon/), ($b$) to firmly prove (or disprove)
the origin of the light-echo within a circumstellar nebula, which would
accurately set the distance, by coronographic observations at widely
different wavelengths and epochs and great spatial resolution, ($c$) to
derive an accurate estimate for the interstellar and the circumstellar
amount of reddening and extinction (including accurate reddening estimates
for many field stars spread over a great range of distances along the line
of sight to V838~Mon), ($d$) when the spectrum will be less perturbed by the
wide P-Cyg profiles now in place and it will better resemble a classical
photosphere, to perform an accurate chemical abundance analysis as a clue to
the outburst causes (like a born-again-AGB scenario and the dredged-up
material it implies).

We conclude this {\sl Letter} with brief considerations about 
possible scenario interpretations for V838~Mon and its outburst that will
be refined in follow-up investigations.

\underline{\sl A classical nova ?} The main similarity resides in the 
rapid rise from quiescence to the first maximum, while several
counter-arguments hold strong, such as the lack of variability and emission
lines from the probable cataclysmic variable precursor. Also the progenitor
spectral energy distribution (an under-luminous F main sequence star) is
quite strange for a CV, and the ejection velocities (not exceeding 500
km~sec$^{-1}$) are low for a classical nova. The very slow evolution could
be interpreted within a classical nova framework as an indication for a
small mass of the accreting WD, and in turn a small mass of the progenitor,
which is however in contrast with the indication of a partnership with the
young galactic disk population.

\underline{\sl A born-again AGB ?} Post-AGB stars on their leftward motion 
on the HR diagram to become the central stars of planetary nebulae can
experience a last helium flash that pushes them back toward the region
occupied by AGB stars. Known examples are FG~Sge, V605~Aql and Sakurai's
object. This is an attractive scenario because V838~Mon in outburst looks
much like a cool AGB with a surface rich in dredged-up barium, lithium,
$s-$elements, and the circumstellar material producing the light-echo agrees
with a recent phase of heavy mass loss at the tip of the AGB (it did not
show up as a PN in quiescence - as it was the case for FG~Sge, V605~Aql and
Sakurai's object - because the precursor was not hot enough to ionize it).
However, the progenitor appears severely under-luminous for the standard
theoretical tracks of post-AGB stars on the HR diagram 
(cf. Bl\"{o}cker and Sch\"{o}nberner 1997), and the rise to maximum is too
fast in comparison with the evolution of outburst of known cases.

\underline{\sl A M31-RedVar or V4332 Sgr analogue ?} In 1989 an erupting star
in the Andromeda Galaxy (M31) developed a M-type cool supergiant spectrum at
maximum, with pronounced P-Cyg profiles and Balmer lines in emission and
peaked to $M_V = -9.95$ (Rich et al. 1989, Mould et al. 1990). The
progenitor was too faint to be identified and the event has been modeled by
Iben and Tutukov (1992) in terms of a cool WD accreting at a very low rate
from a companion and under such circumstances the entire WD could experience
a thermonuclear runaway. A similar type of eruption brought V4332~Sgr in
1994 from anonymous quiescence at $V\sim17$ to $V\sim$8.4~mag at maximum,
developping an M-giant spectrum that progressed toward later spectral types
along the outburst evolution, with significant emission in the Balmer lines
and [OI], [FeII], FeI, MgI and NaI but without marked P-Cyg type profiles
(Martini et al. 1999). V838~Mon shares common characteristics with both such
events, but some differences remain. First, both M31-RedVar and
V4332~Sgr had single-peaked and smooth light-curves, while V838~Mon
experienced three maxima of widely different photometric and spectroscopic
characteristics. The emission limited to only the Balmer lines resembles
M31-RedVar but the $M_V = -4.35$ peak brightness is much closer to V4332~Sgr.
The presence and variability of P-Cyg profiles in V838~Mon match
what saw in M31-RedVar, while the $\sim$F dwarf progenitor is
similar to the $\sim$K dwarf visible progenitor of V4332~Sgr. It could be that these 
three events are manifestations of the same and new class of astronomical 
objects.

\end{document}